\documentclass[conference]{IEEEtran}
\IEEEoverridecommandlockouts
\usepackage{cite}
\usepackage{amsmath,amssymb,amsfonts}
\usepackage{algorithmic}
\usepackage{graphicx}
\usepackage{textcomp}
\usepackage{xcolor}

\usepackage{booktabs} 
\usepackage{xfrac, amsthm, tabularx, subcaption}

\usepackage{mathtools}
\usepackage{alltt, xspace, times, epsfig, url}
\usepackage{gensymb,array,bm}

\DeclareMathOperator*{\argmin}{\arg\!\min}

\def\BibTeX{{\rm B\kern-.05em{\sc i\kern-.025em b}\kern-.08em
    T\kern-.1667em\lower.7ex\hbox{E}\kern-.125emX}}

\begin{document}

\title{Utility-Optimized Synthesis of Differentially \\Private Location Traces}

\author{\IEEEauthorblockN{M.~Emre Gursoy}
\IEEEauthorblockA{\textit{Department of Computer Engineering} \\
\textit{Ko\c{c} University}\\
Istanbul, Turkey \\
emregursoy@ku.edu.tr}
\and
\IEEEauthorblockN{Vivekanand Rajasekar}
\IEEEauthorblockA{\textit{School of Computer Science} \\
\textit{Georgia Institute of Technology}\\
Atlanta, GA, USA \\
vivekraja07@gmail.com}
\and
\IEEEauthorblockN{Ling Liu}
\IEEEauthorblockA{\textit{School of Computer Science} \\
\textit{Georgia Institute of Technology}\\
Atlanta, GA, USA \\
ling.liu@cc.gatech.edu}
}

\maketitle

\begin{abstract}
Differentially private location trace synthesis (DPLTS) has recently emerged as a solution to protect mobile users' privacy while enabling the analysis and sharing of their location traces. A key challenge in DPLTS is to best preserve the utility in location trace datasets, which is non-trivial considering the high dimensionality, complexity and heterogeneity of datasets, as well as the diverse types and notions of utility.
In this paper, we present OptaTrace: a utility-optimized and targeted approach to DPLTS. Given a real trace dataset $D$, the differential privacy parameter $\varepsilon$ controlling the strength of privacy protection, and the utility/error metric $Err$ of interest; OptaTrace uses Bayesian optimization to optimize DPLTS such that the output error (measured in terms of given metric $Err$) is minimized while $\varepsilon$-differential privacy is satisfied. In addition, OptaTrace introduces a utility module that contains several built-in error metrics for utility benchmarking and for choosing $Err$, as well as a front-end web interface for accessible and interactive DPLTS service. Experiments show that OptaTrace's optimized output can yield substantial utility improvement and error reduction compared to previous work. 
\end{abstract}



\section{Introduction}

As mobile devices and location-based services become increasingly ubiquitous, there is growing interest in analyzing and sharing information derived from mobile users' location traces. For example, Uber Movement shares anonymized data aggregated from billions of trips to help urban planning around the world \cite{UberMovement}. Google's COVID-19 Community Mobility Reports  share insights regarding movement trends over time by category (retail, grocery stores, pharmacies, transit stations, and so forth), which are also used in products such as Google Maps \cite{GoogleCOVID19}. NYC Taxi and Limousine Commission shares taxi ride logs from New York City. Yet, the highly sensitive nature of mobile users' location traces gives rise to privacy risks when analyzing or sharing location data. Recent research has shown that many privacy attacks remain relevant despite aggregation or anonymization, such as stalking, trajectory reconstruction, de-anonymization, and membership inference attacks \cite{xu2017trajectory,pyrgelis18knock,ma2013privacy,wang2018anonymization,chang2018revealing,kaplan2018location,pyrgelis2020measuring}. 

Differentially private location trace synthesis (DPLTS) has emerged as a solution to protecting mobile users' privacy while analyzing and sharing information derived from their traces \cite{gursoy2018differentially,gursoy2018utility,he2015dpt,chen2012stm,chen2012differentially}. In DPLTS, a generative synthesis system takes as input the dataset consisting of mobile users' real location traces (denoted $D$) and outputs a synthetic location trace dataset (denoted $D_{syn}$) which is syntactically and semantically similar to $D$, but consists of traces built while satisfying differential privacy. $D_{syn}$ can then be used for in-house data analytics or for public release of statistics. DPLTS has two main privacy benefits. First, differential privacy provides a formal and robust privacy guarantee such that $D_{syn}$ does not reveal the presence, absence or content of any real trace in $D$. Second, since the traces in $D_{syn}$ are synthetic, they do not have one-to-one correspondence with any real individual; thus, re-identification and record linkage attacks are thwarted. 

A central challenge in DPLTS, however, is how to best preserve the utility and statistical characteristics of $D$ when synthesizing $D_{syn}$. This is a non-trivial challenge, considering the high dimensionality, complexity and heterogeneity of location trace datasets, e.g., varying dataset cardinality, trace length, trace duration, density, and sampling rate. In addition, there are endlessly many applications and statistics that could be derived from $D_{syn}$, such as travel time estimation, spatial density extraction and mobility pattern mining. Given that a different error metric or utility metric would be appropriate for each task, it is not feasible that a static DPLTS method preserves \textit{all} utilities simultaneously.

Motivated by the above, this paper studies the following problem. Given a real trace dataset $D$, the differential privacy budget $\varepsilon$ controlling the strength of privacy protection, and the utility/error metric $Err$ of interest, we wish to optimize DPLTS such that output $D_{syn}$ minimizes $Err$ between $D$ and $D_{syn}$ while satisfying $\varepsilon$-differential privacy. Towards this goal, we design and develop the OptaTrace system which extends the AdaTrace system \cite{gursoy2018utility}. OptaTrace uses Bayesian optimization, a black-box optimization method, to find optimized parameters and budget distributions for AdaTrace's synopsis module which minimize error according to the given $D$, $\varepsilon$ and $Err$. Furthermore, contributions of OptaTrace also include: (i) a utility module which contains several built-in error metrics to choose $Err$, as well as allowing the specification of a novel $Err$ metric; and (ii) a front-end web interface for user-friendly and interactive DPLTS service. The user can upload their $D$, choose $\varepsilon$ and $Err$ through the web interface, as well as visually explore statistics regarding output $D_{syn}$ or download $D_{syn}$ to their local machine for further analysis.


OptaTrace provides a utility-targeted approach: If the utility metric $Err$ is known ahead of time or can be approximated, OptaTrace's output $D_{syn}$ can yield substantial utility improvement compared to untargeted (non-optimized) DPLTS approaches. We experimentally demonstrate the utility improvement of OptaTrace using three datasets, three $\varepsilon$ values and four error metrics. Compared to the state-of-the-art AdaTrace system, OptaTrace outperforms AdaTrace in all experiments, and provides up to 50\% reduction in utility loss. Our experiments also show that the optimized parameters are different for different $D$, $\varepsilon$ and $Err$; which demonstrates the necessity of individual case-by-case optimization for targeted utility improvement.

The rest of this paper is organized as follows. In Section \ref{sec:Background}, we review the location trace data model and differential privacy background. In Section \ref{sec:System}, we describe the OptaTrace system design. In Section \ref{sec:Implementation}, we give the implementation details of OptaTrace as well as a brief demonstration of its front-end web interface. Section \ref{sec:Experiments} provides the results of our experimental evaluation. We summarize related work in Section \ref{sec:RelatedWork} and conclude in Section \ref{sec:Conclusion}.

\section{Data Model and Privacy Background} \label{sec:Background}

Consider a dataset $D = \{T_1, T_2, ..., T_{|D|} \}$ where each $T_i$ corresponds to one mobile user's location trace. In order to protect the privacy of users' location traces, we enforce the popular notion of differential privacy \cite{dwork2008differential,dwork2014algorithmic} as follows. Let $nbrs(D)$ denote the set of datasets neighboring $D$, such that for all $D' \in nbrs(D)$ the following holds: $(D - D') \cup (D' - D) = \{T \}$ where $T$ denotes one location trace. Then, we say that a randomized algorithm $\mathcal{A}$ satisfies $\varepsilon$-differential privacy ($\varepsilon$-DP) if for all datasets $D$ and $D' \in nbrs(D)$ and for all outcomes of the algorithm $S \in \textit{Range}(\mathcal{A})$: 
\[
\frac{\text{Pr}[\mathcal{A}(D) = S]}{\text{Pr}[\mathcal{A}(D') = S]} \leq e^{\varepsilon}
\]
Here, $\varepsilon$ is called the privacy budget, which determines the strength of privacy protection. Smaller $\varepsilon$ gives stronger privacy. 

Note that the above is a trace-level enforcement of differential privacy, i.e., it asserts that the outcome of the algorithm $\mathcal{A}$ will not enable an adversary to distinguish, beyond a probability controlled by $\varepsilon$, between two datasets $D$ and $D'$ that differ by a complete location trace $T$. This protects the complete location trace of a mobile user, and differs from DP perturbation of individual location points when the user is querying a location-based service \cite{andres2013geo,yu2017dynamic}. 

Differential privacy has three properties which are relevant and useful in the design of OptaTrace:
\begin{itemize}
    \item \textit{Sequential Composition}: For $n$ algorithms $\mathcal{A}_1 \ldots \mathcal{A}_n$ each satisfying DP with budget $\varepsilon_1 \ldots \varepsilon_n$, the sequential execution of these algorithms on $D$ satisfies ($\sum_{i=1}^{n}\varepsilon_i$)-DP. 
    \item \textit{Parallel Composition}: For two algorithms $\mathcal{A}_1$ and $\mathcal{A}_2$ satisfying $\varepsilon_1$-DP and $\varepsilon_2$-DP respectively, if $\mathcal{A}_1$ and $\mathcal{A}_2$ are executed on disjoint subsets of $D$, the resulting execution satisfies max$(\varepsilon_1, \varepsilon_2 )$-DP.
    \item \textit{Immunity to Post-Processing}: Let $S$ denote the outcome of an $\varepsilon$-DP algorithm $\mathcal{A}$ executed on $D$, i.e., $\mathcal{A}(D)=S$. Then, any post-processing of $S$, including its use in a future algorithm or its public release, does not violate the $\varepsilon$-DP guarantee of $S$. 
\end{itemize}

\section{OptaTrace System} \label{sec:System}

\begin{figure*}
    \centering
    \includegraphics[width=0.7\textwidth]{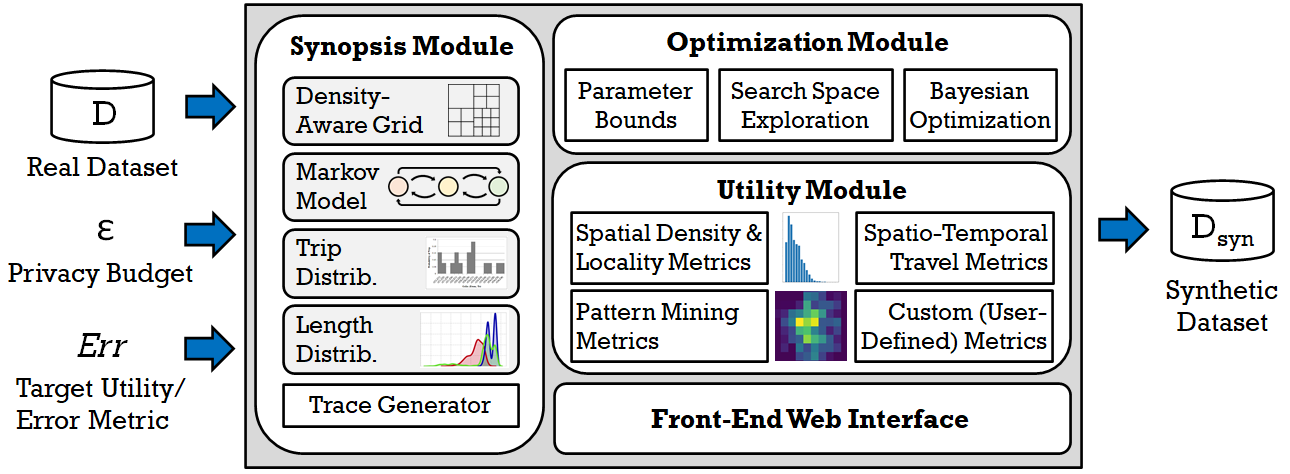} 
    \caption{OptaTrace system architecture}
    \label{fig:arch}
    \vspace{-8pt}
\end{figure*}



The goal of our OptaTrace system can be stated as follows: Given a real dataset $D$ of actual location traces, the differential privacy budget $\varepsilon$, and the target utility/error metric $Err$, generate a synthetic location trace dataset $D_{syn}$ such that $\varepsilon$-DP is satisfied and the utility loss between $D$ and $D_{syn}$ measured in terms of $Err$ is minimized. 

To achieve this goal, we designed the OptaTrace system as shown in Figure \ref{fig:arch}. It consists of four modules: synopsis mod\-ule, optimization module, utility module and front-end web interface. In this section, we explain each module one by one.

OptaTrace extends the state of the art AdaTrace system \cite{gursoy2018utility} in three ways. First, OptaTrace includes a Bayesian optimization module for optimizing the parameter distribution according to given $D$, $\varepsilon$ and $Err$. The optimization module iteratively searches for the optimized parameters that minimize $Err$, which are often different for different $D$, $\varepsilon$ or $Err$. Second, OptaTrace includes a utility module which contains four categories of error metrics, so that the OptaTrace user can choose $Err$ from existing metric categories or implement a new $Err$ metric. The utility module of OptaTrace can also be used for benchmarking and evaluation of different $D$ and $D_{syn}$. Third, OptaTrace provides a front-end web interface which enables OptaTrace users to seamlessly upload their $D$, choose their desired privacy level $\varepsilon$ and metric $Err$ through their favorite web browser. Preliminary statistics regarding the output $D_{syn}$ can be obtained through OptaTrace's web interface, and $D_{syn}$ can also be downloaded for further analysis. 

\subsection{Synopsis Module} \label{sec:SynopsisModule}

The synopsis module of OptaTrace contains four features for extracting useful statistical information from $D$ while satisfying differential privacy: density-aware grid $\mathbb{A}$, Markov model $\mathcal{M}$, trip distribution $\mathcal{R}$ and length distribution $\mathcal{L}$. These four features are then used by the trace generator (fifth component of the synopsis module) to generate synthetic traces which are added to $D_{syn}$. Below, we give brief descriptions of the four features and the trace generator. Full technical descriptions and privacy proofs can be found in \cite{gursoy2018differentially,gursoy2018utility}. 

In order to satisfy $\varepsilon$-DP as a whole when extracting four features, OptaTrace makes use of DP's composition and post-processing properties. In particular, extracting the density-aware grid satisfies $(w_1 \times \varepsilon)$-DP, the Markov model satisfies $(w_2 \times \varepsilon)$-DP, the trip distribution satisfies $(w_3 \times \varepsilon)$-DP, and the length distribution satisfies $(w_4 \times \varepsilon)$-DP where the sum of the weights is: $\sum^4_{i=1} w_i = 1$. Thus, by sequential composition, the total of the four features satisfy $\varepsilon$-DP. The trace generator only uses the four features without modifying them or accessing the real dataset $D$, therefore $\varepsilon$-DP still holds due to immunity to post-processing.

\textbf{Density-Aware Grid $\mathbb{A}$:} Accurately encoding the location space of $D$ is the first step towards extracting useful statistics from $D$. We use a 2-dimensional grid structure to encode the location space of $D$, which is a common encoding strategy for location data. Yet, choosing an appropriate grid size and structure is non-trivial under DP and efficiency constraints. If the grid is too coarse (3x3), then each grid cell covers a large spatial area, and knowing that $T$ visited a certain cell is uninformative. If the grid is too detailed (50x50), then there arise many empty cells with zero density, but noise must still be added to each of these cells to satisfy DP, which causes DP noise to overwhelm useful statistics, and inefficiency due to a large number of redundant empty cells.

In order to find a good balance, OptaTrace uses a density-aware grid structure $\mathbb{A}$ which adapts the number of cells that cover a geographic region according to the \textit{density} of the region, i.e., the number of location readings in $D$ that originate from that region. For low density regions, $\mathbb{A}$ places few large cells. For high density regions, $\mathbb{A}$ divides the region into many small cells. $\mathbb{A}$ is constructed in three steps: (1) Initially, an $N \times N$ uniform grid is laid in the geographic space covered by $D$, resulting in a total of $N^2$ cells. (2) For each cell, a density query is issued on $D$ to retrieve how many normalized location readings exist in that cell. The answer to each density query is perturbed with randomized noise to satisfy DP. (3) Depending on their density, each of the original $N^2$ cells is either kept as is, or divided internally into smaller cells. Higher density implies more division, e.g., an extremely dense cell may be divided further into $6 \times 6$ smaller cells, whereas a medium density cell may be divided further into $2 \times 2$ smaller cells. The resulting grid by the end of step 3 is denoted $\mathbb{A}$.

\begin{figure*}
\centering
\begin{minipage}[b]{0.20\textwidth}
\includegraphics[width=\textwidth]{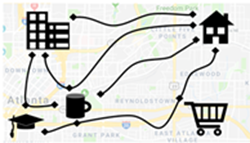}
\subcaption{}
\label{fig:traces}
\end{minipage}%
\hspace{3mm}
\begin{minipage}[b]{0.13\textwidth}
\includegraphics[width=\textwidth]{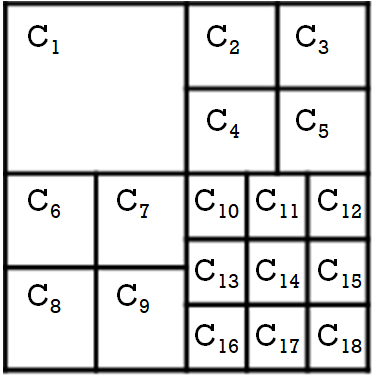}
\subcaption{}
\label{fig:grid}
\end{minipage}%
\hspace{3mm}
\begin{minipage}[b]{0.20\textwidth}
\includegraphics[width=\textwidth]{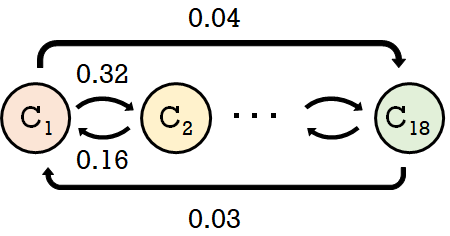}
\subcaption{}
\label{fig:markov}
\end{minipage}%
\hspace{3mm}
\begin{minipage}[b]{0.20\textwidth}
\includegraphics[width=\textwidth]{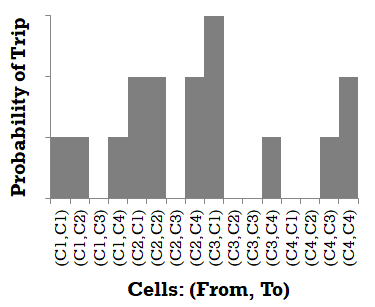}
\subcaption{}
\label{fig:trip}
\end{minipage}%
\hspace{3mm}
\begin{minipage}[b]{0.18\textwidth}
\includegraphics[width=\textwidth]{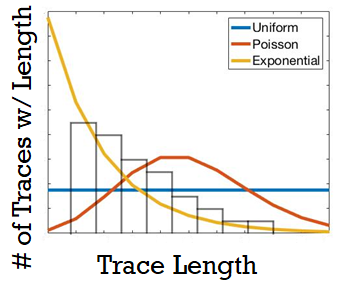}
\subcaption{}
\label{fig:length}
\end{minipage}%
\caption{Visualization of sample location traces and components of the synopsis module. (a) Visualization of real location traces. (b) Adaptive grid $\mathbb{A}$ with cells numbered $C_1$ to $C_{18}$. (c) Markov mobility model $\mathcal{M}$. (d) Trip distribution $\mathcal{R}$ (partially shown). (e) Calculation of length distribution $\mathcal{L}$ for a pair of cells $C_{a} \leadsto C_b$.}
\label{fig:synopsisModule}
\vspace{-8pt}
\end{figure*}

An example density-aware grid $\mathbb{A}$ is given in Figure \ref{fig:grid}. A $2 \times 2$ grid was initialized in step 1. The top-left cell was left without any further division due to low density. The top-right and bottom-left cells were each divided further into $2 \times 2$ cells due to having medium density. The bottom-right cell was divided into $3 \times 3$ cells due to having high density.


\textbf{Markov Model $\mathcal{M}$:} OptaTrace employs a Markov chain to model intra-trace mobility and movement behavior. Markov chains are a popular technique for mobility modeling, with many works showing that they are accurate in capturing urban mobility in real datasets and predicting users' next locations \cite{gambs2012next,lu2013approaching,rathore2019scalable}. Our Markov model, denoted $\mathcal{M}$, contains:
\begin{itemize}
    \item A set of Markov states: Each state corresponds to a cell $C$ from the grid $\mathbb{A}$. 
    \item Transition probabilities between states: For each pair of states $C_i$ and $C_j$, there exists a transition probability for moving from state $C_i$ to state $C_j$. 
\end{itemize}
The transition probabilities are learned from the input real trace dataset $D$. Calibrated noise is added to each transition probability to satisfy DP. A sample Markov model is visualized in Figure \ref{fig:markov}, where each state corresponds to a cell from grid $\mathbb{A}$, and the transition probabilities are written next to the transition arrows between each state.

\textbf{Trip Distribution $\mathcal{R}$:} A mobile user's movement throughout the day often consists of several trips, e.g., home-work commute, lunch trip to a restaurant, trip to the gym after work, and so forth. Furthermore, real-life location trace datasets such as taxi or Uber traces often consist of a collection of trips. The trip distribution $\mathcal{R}$ in OptaTrace aims to preserve the joint association between the start-end locations of trips, which is useful for tasks including passenger demand analysis, taxi destination prediction, city planning, and so forth. 

Let $T: C_i \leadsto C_j$ denote that location trace $T$ starts its trip in cell $C_i$ and finishes its trip in cell $C_j$. In essence, the trip distribution $\mathcal{R}$ is a probability mass function that contains one probability entry for each pair of cells $(C_i, C_j) \in \mathbb{A} \times \mathbb{A}$ that captures what percentage of traces in $D$ make the trip $C_i \leadsto C_j$. Let $D_{C_i \leadsto C_j}$ denote the subset of $D$ which consists of traces that make the trip $C_i \leadsto C_j$. Trip distribution $\mathcal{R}$ is:
\begin{equation}
\mathcal{R}_{D,\mathbb{A}} ((C_i,C_j)) := 
\begin{cases}
\frac{|D_{C_i \leadsto C_j}|}{|D|} & \text{ for } (C_i,C_j) \in \mathbb{A} \times \mathbb{A} \\
0 & \text{ otherwise}
\end{cases}
\end{equation}
Cardinalities $|D_{C_i \leadsto C_j}|$ and $|D|$ are perturbed with noise to satisfy DP. An example trip distribution is visualized in Figure \ref{fig:trip}. (Note that this is a partial figure containing only 16 entries on the x-axis because of the space constraint. The actual trip distribution contains $\mathbb{A} \times \mathbb{A}$ entries.)

\textbf{Length Distribution $\mathcal{L}$:} There are likely to be multiple trips between a pair of cells $C_a \leadsto C_b$ and they may have varying length. OptaTrace learns the statistical length distribution for trips between $C_a \leadsto C_b$ as follows. First, the length of each trace $T \in D_{C_{a} \leadsto C_{b}}$ is measured. 
Second, a histogram is built based on how many traces in $D_{C_{a} \leadsto C_{b}}$ have each length, e.g., 10 traces have length 2, 5 traces have length 3, and so forth. Third, three statistical distributions (Uniform, Poisson and Exponential) are initialized as potential candidates to represent the observed histogram. Finally, a goodness of fit test is used to determine which distribution best fits the observed histogram. The best fit distribution is stored in $\mathcal{L}$ and the rest are discarded. DP is enforced during the process of building the candidate distributions (one of which is eventually stored in $\mathcal{L}$) by adding noise to the parameters of the distributions.

We visualize the computation of the length distribution for one choice of $C_a \leadsto C_b$ in Figure \ref{fig:length}. The true length histogram is shown in bars. The three statistical distributions initialized while satisfying DP (Uniform, Poisson and Exponential) are shown with different colored lines. In this particular example, the goodness of fit test selects the Exponential distribution as the best fit, since its shape is closest to the shape of the histogram. Thus, the Exponential distribution would be stored in the length distribution for $C_a \leadsto C_b$.

\textbf{Trace Generator:} The trace generator is a synthesis algorithm which takes as input the previously computed four elements of the synopsis (density-aware grid $\mathbb{A}$, Markov model $\mathcal{M}$, trip distribution $\mathcal{R}$ and length distribution $\mathcal{L}$) and outputs a synthetic dataset of traces denoted $D_{syn}$ with number of traces equal to cardinality of $D$. The trace generator does not modify the four existing synopsis elements or access the real dataset $D$. Since the synopsis elements already satisfy $\varepsilon$-DP as a whole, and since the execution of the trace generator performs only sampling and post-processing on the synopsis elements, the $\varepsilon$-DP guarantee still holds. 

The trace generator generates each synthetic trace one by one, and adds them to $D_{syn}$ upon generation. The steps to generate one synthetic trace denoted $T_{syn}$ are as follows:
\begin{enumerate}
    \item Draw a sample from $\mathcal{R}$ to determine the trip for $T_{syn}$. Let $(C_{start}, C_{end})$ denote the sampled trip.
    \item Draw a sample from $\mathcal{L}$ to determine the length of $T_{syn}$. Let $\ell$ denote the sampled length.
    \item Initialize $T_{syn}$ with length $\ell$, starting cell equal to $C_{start}$, and end cell equal to $C_{end}$. 
    \item To determine each of the intermediate locations in $T_{syn}$, perform a random walk on Markov chain $\mathcal{M}$. (Random walk is guaranteed to start in $C_{start}$ and end in $C_{end}$.) 
\end{enumerate}
This process results in a synthetic trace consisting of exactly one trip, which is suitable when $D$ or $D_{syn}$ consists of Uber trips or taxi trips. Longer location traces (e.g., a mobile user's trace for one day or longer) are likely to contain multiple consecutive trips. In such situations, we extend the above process such that the $C_{end}$ of the previous trip becomes the $C_{start}$ of the next trip.

\subsection{Utility Module} \label{sec:UtilityModule}

Recall that $D$ denotes the input real dataset and $D_{syn}$ denotes the output synthetic dataset. It is desired that $D_{syn}$ preserves as much utility and statistical similarity to $D$ as possible while satisfying the $\varepsilon$-DP guarantee. However, due to the noise addition in OptaTrace to satisfy $\varepsilon$-DP, $D_{syn}$ will incur some utility loss compared to $D$. The goal of the utility module is to provide metrics for utility loss measurement.

Since location traces are inherently complex and utility in the location data analytics domain is a multi-faceted concept, there are many ways in which utility loss can be measured. Also, utility loss often depends on the end application and how $D_{syn}$ will be used by the data analyst. For example, if $D_{syn}$ will be used for building population density heatmaps, accurate representation of the location space and density preservation of $D$ will be most important. In contrast, if $D_{syn}$ will be used for analyzing taxi/Uber passenger demand, then preserving trip distributions will be most important. Consequently, the utility module of OptaTrace is designed to include a diverse set of built-in metrics for utility loss measurement, and also be extensible so that new metrics can be added in the future. Metrics in the utility module can be presented under 4 categories:

\textbf{Spatial Density and Locality Metrics:} Several geospatial analytics tasks rely on spatial densities and localities, such as Point-of-Interest analysis, spatial heatmaps, and location-based advertisement. 
Google's COVID-19 Community Mobility Reports \cite{GoogleCOVID19} is a recent example requiring the preservation of spatial densities: each report highlights the percentage change in visits to places such as grocery stores, restaurants, parks and transit stations in a city when compared to a regular day before COVID-19. Utility metrics that measure error between $D$ and $D_{syn}$ in terms of spatial density and locality include: (i) error in computing the number of visits to a location using $D_{syn}$ versus using $D$, (ii) error in determining location popularity rankings using $D_{syn}$ versus $D$, e.g., error in restaurant popularity rankings, (iii) error in computing answers to a range query workload using $D_{syn}$ versus $D$, and so forth.

\textbf{Spatio-Temporal Travel Metrics:} Since location trace datasets often consist of taxi/Uber rides or daily commutes, analyzing aggregate trip features may yield not only a commercial advantage but also an urban planning advantage. For example, Uber Movement \cite{UberMovement} provides a web interface for calculating average travel times between different neighborhoods, average road speeds at different times of day, etc. 
These statistics may be computed using $D_{syn}$ rather than $D$ to enforce formal privacy protection. Utility metrics that are suitable in measuring error in such an approach include: (i) error in computing average number of daily trips between two neighborhoods using $D_{syn}$ versus $D$, (ii) error in estimating average street speed using $D_{syn}$ versus $D$, (ii) error in estimating travel time between two neighborhoods using synthetic trip data in $D_{syn}$ versus actual historical trip data in $D$, etc.

\textbf{Pattern Mining Metrics:} Pattern mining and pattern retrieval have been critical research problems in trajectory data mining. They have applications to not only human mobility patterns in urban environments but also to wildlife animals, e.g., finding seasonal migration patterns. Let $\mathcal{P}$ denote the results of pattern mining on $D$ and $\mathcal{P}_{syn}$ denote the results of pattern mining on $D_{syn}$. The error between $\mathcal{P}$ and $\mathcal{P}_{syn}$ is measured by metrics including: (i) the set similarity between $\mathcal{P}$ and $\mathcal{P}_{syn}$, e.g., Jaccard similarity and F1 score, and (ii) the observation frequency of a pattern in $\mathcal{P}$ versus $\mathcal{P}_{syn}$.

\textbf{Custom (User-Defined) Metrics:} There can be metrics that are not covered by the categories and applications above. The design of OptaTrace allows the OptaTrace user to implement new, custom error metrics. The new metric can be a (weighted) combination of existing metrics, as well as a completely new metric inspired by a novel use case or unforeseen application of a location trace dataset. 




\subsection{Optimization Module} \label{sec:OptimizationModule}

Recall from Section \ref{sec:SynopsisModule} that $(w_1, w_2, w_3, w_4)$ are the four weight parameters of the OptaTrace system, and let $Err$ denote the target error metric that is sought to be minimized. The goal of the optimization module can be stated as finding the values of $w_1, w_2, w_3, w_4$ such that:
\begin{equation}
    \argmin_{w_1, w_2, w_3, w_4} ~ Err(D, D_{syn})
\end{equation}
That is, the optimization module aims to find the set of parameters for OptaTrace such that the output $D_{syn}$ has lowest amount of error possible while satisfying $\varepsilon$-DP. 

In order to achieve this goal, the optimization module uses \textit{Bayesian optimization}, which is a class of machine learning-based methods for black-box function optimization \cite{brochu2010tutorial,snoek2012practical}. Its strategy is to treat the behavior of the synopsis module as a black-box function that needs to be optimized. First, it places a random prior regarding how the function behaves. Then, it gathers several evaluations of the function, e.g., executions with different parameter values $(w_1, w_2, w_3, w_4)$ under the given $D$ and $\varepsilon$. After observing the output of the function, i.e., the resulting error with the given set of parameters, it updates its belief regarding function behavior. Next, in each iteration the set of parameters is selected according to past observations and updated belief regarding which direction is best to explore for minimizing error. After several iterations, parameters converge to their optimized values which minimize $Err(D, D_{syn})$ under the given $\varepsilon$ and $D$. Our use of Bayesian optimization can be explained in three consecutive steps. 


\textbf{(1) Specification of Parameter Bounds:} In order to cons\-train the optimization search space, the initial step is to specify the bounds of each parameter that needs to be optimized. We specify the following constraints, which ensure that $\varepsilon$-DP is satisfied as a whole. 
\begin{equation}
    0 < w_1, w_2, w_3, w_4 < 1 \text{~~~and~~~} \sum^4_{i=1} w_i = 1
\end{equation}

\textbf{(2) Search Space Exploration:} Given the parameter constr\-aints and the target error metric $Err$, we perform several \textit{random explorations} to explore the search space, by executing the synopsis module with different random parameter sets and observing the resulting errors. This helps diversify the optimizer's prior beliefs and ensures that the search space is sufficiently probed before the actual optimization process begins. By default, we use 100 explorations.

\textbf{(3) Iterative Bayesian Optimization:} We execute several iterations of Bayesian optimization (by default, 100 iterations). In each iteration, the optimizer selects a set of parameters $(w_1, w_2, w_3, w_4)$, executes the synopsis module, and observes the resulting error $Err$ in terms of the given error metric. The observed error is used to update the optimizer's belief and informs the choice of parameters in the next iteration. As the number of observations grows, the optimizer becomes more certain which regions in the parameter space are more worth exploring. In time, the parameters converge to their optimized values which minimize $Err$. 

\subsection{Front-End Web Interface} \label{sec:Interface}

We designed a front-end web interface for OptaTrace so that OptaTrace users can access and interact with OptaTrace through a user-friendly and interactive web interface. Our goals in designing the front-end web interface include:
\begin{itemize}
    \item Accessible and easy-to-use privacy functionality for non-experts: Data analysts may be interested in adopting a privacy technology when performing location data analytics. However, off-the-shelf differential privacy (DP) tools are scarce, and they are often difficult to use for those who are not experts in DP. OptaTrace’s web interface addresses this problem by providing an easy-to-use and interactive DP enforcement opportunity to data analysts. 
    \item Differential privacy-as-a-service: Laws and regulations (such as GDPR) are increasingly restricting the storage of raw, sensitive user data. As a result, companies and businesses are turning towards innovative privacy protection methods so that user data can be privatized before being used or stored. OptaTrace’s web interface can provide data privatization service in the following manner. Consider that a company collects mobile users’ location trace data, collectively denoted by $D$, as in Figure 1. Using the web interface, $D$ is input to OptaTrace along with the privacy budget $\varepsilon$. The output dataset $D_{syn}$ is downloaded and stored by the company; and the real data $D$ is destroyed afterwards. As such, OptaTrace can serve as a differential privacy-as-a-service tool.
    \item Extensibility for client-server use and web hosting: The front-end design is suitable for client-server environments such that the OptaTrace software runs on a server machine, clients remotely connect to the server through the Internet, and they benefit from the OptaTrace privacy service. While we developed and tested the front-end web interface primarily in a single client environment, the client-server functionality may be extended in the future to enable OptaTrace be hosted on a central powerful web server, and clients use the OptaTrace service by connecting via their web browser.
\end{itemize}

\section{OptaTrace System Implementation} \label{sec:Implementation}

\subsection{Implementation Details}

The implementation of OptaTrace consists of three main parts: Web UI server, Python component, and Java component.

\textbf{Web UI server} provides the client-facing front-end web interface. It runs on Vue.js, a progressive open-source UI framework. It uses Vuetify, a Material Design component framework that provides a modern look, and Axios, an HTTP client for the browser. The Web UI server is used by the client to upload dataset $D$, choose privacy and optimization parameters, and download/analyze the output dataset $D_{syn}$. Axios is used to make REST API calls over HTTP to communicate the client's choices with the Python component. When the Web UI server needs to display information, it makes an HTTP request to the Python component, which in turn makes calls to the Java component.

\textbf{Python component} is written in Python and contains the optimization module of OptaTrace. It sits between the Web UI server and the Java component. It communicates with the Web UI server using REST API calls over HTTP that are handled using the Flask library. Upon receiving clients' commands and choices from the Web UI server, the Python component uses the Bayesian optimization library to iteratively perform parameter optimization. Each iteration requires back-and-forth communication with the Java component.

\textbf{Java component} contains the synopsis module and utility module of OptaTrace, written in Java language. It communicates with the Python component by using the Py4J library. Py4J enables Python programs running in a Python interpreter to dynamically access Java objects in a Java Virtual Machine \cite{Py4J}. Methods are called as if the Java objects resided in the Python interpreter and Java collections can be accessed through standard Python collection methods. Py4J also enables Java programs to call back Python objects. This allows the Python component to call methods from the Java codebase as if it were simply an extension of the Python component, which enables fast communication and data transfer between the Java component and Python component.

\subsection{Execution and Brief Demonstration}

\begin{figure}
\centering
\begin{minipage}[b]{0.48\textwidth}
\includegraphics[width=\textwidth]{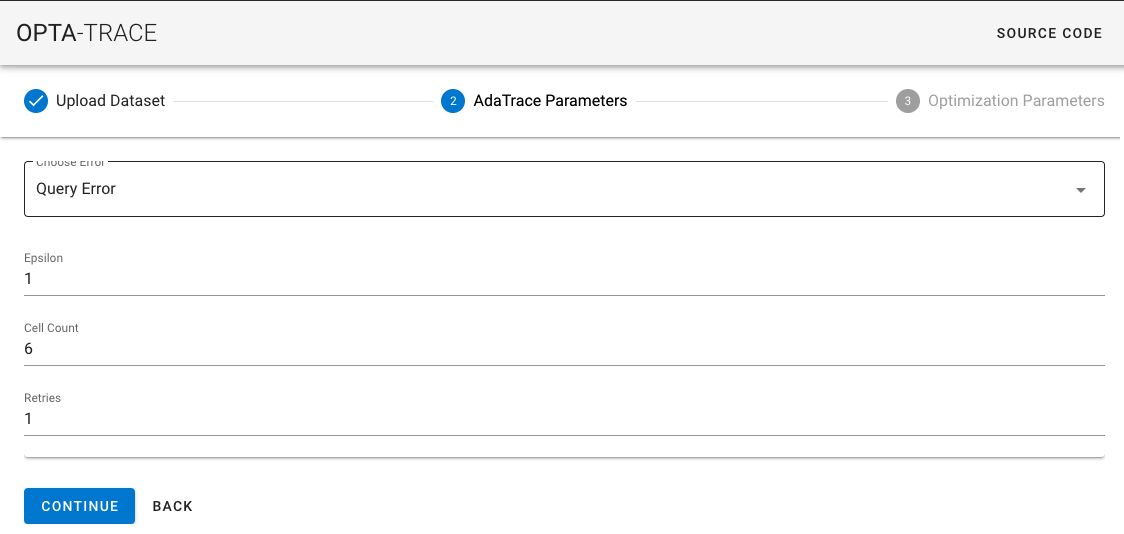}
\subcaption{Input Phase}
\label{fig:inputPhase}
\end{minipage}\\ \vspace{8pt}
\begin{minipage}[b]{0.48\textwidth}
\includegraphics[width=\textwidth]{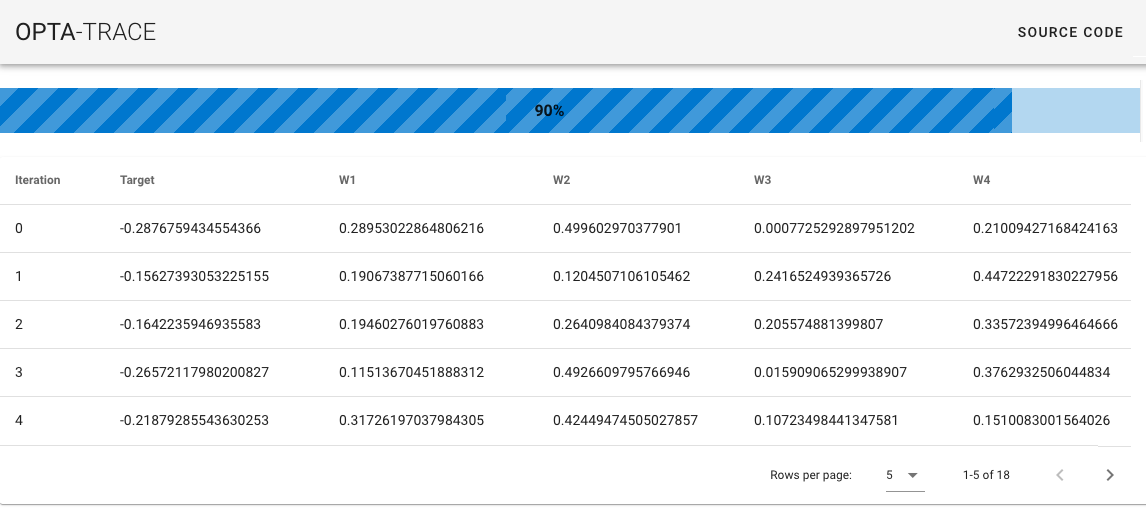}
\subcaption{Computation and Optimization Phase}
\label{fig:optimizationPhase}
\end{minipage}\\ \vspace{8pt}
\begin{minipage}[b]{0.48\textwidth}
\includegraphics[width=\textwidth]{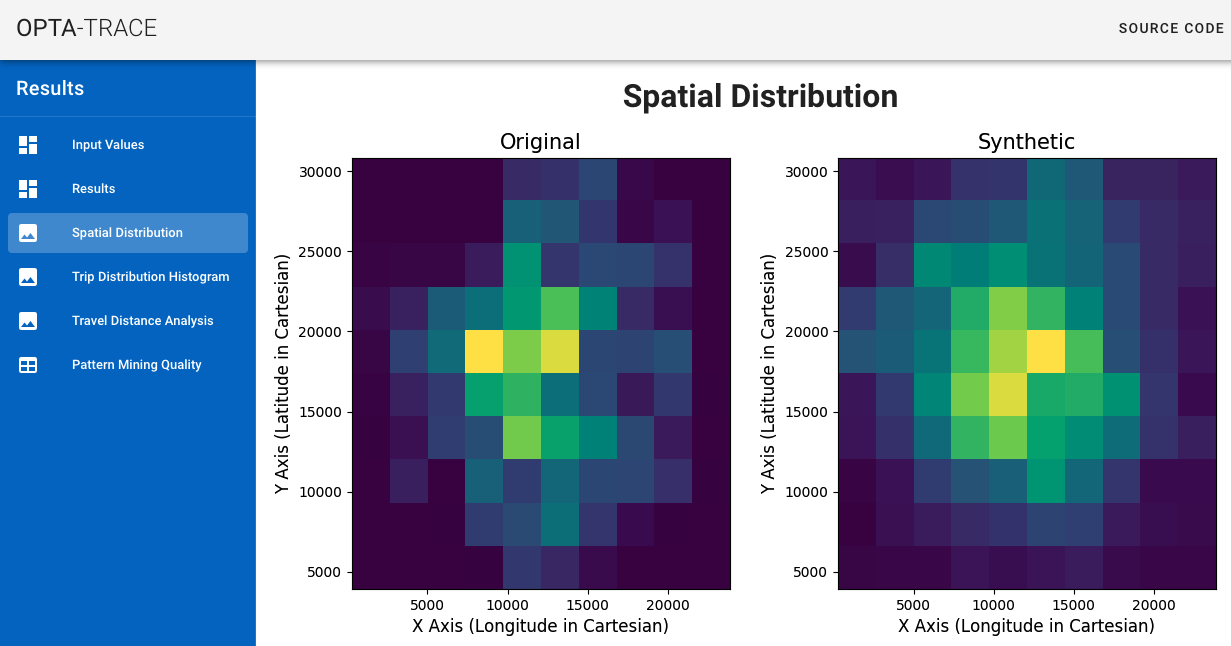}
\subcaption{Results and Analysis Phase}
\label{fig:resultsPhase}
\end{minipage}%
\caption{Screenshots from the OptaTrace front-end web interface showing the three main phases of user interaction.}
\label{fig:phases}
\vspace{-8pt}
\end{figure}

Combining the components listed in the previous section, the execution of OptaTrace has three phases. First, the Web UI server receives $D$, $\varepsilon$ and related parameters from the OptaTrace user. Second, these are sent to the Python component which starts the optimization process. The Python component and the Java component communicate back-and-forth for many iterations of Bayesian optimization. In each iteration, the Python component instructs the Java component to run the synopsis module and utility module with a certain setting of parameters, observes the results, and updates the parameter settings for the next iteration. Third, after the optimization is complete, final results ($D_{syn}$ and related statistics) are computed by the Java component, sent to the Python component, and then forwarded to the Web UI server. They are visualized and displayed to the OptaTrace user through the graphical web interface. 
In this section, we describe the three phases one-by-one and provide a screenshot for each phase in Figure \ref{fig:phases}. 


\textbf{Input Phase:} In this first phase, the OptaTrace user is asked to provide the necessary inputs such as $D$, $\varepsilon$, $Err$ metric. The phase consists of three substeps: Upload Dataset, AdaTrace Parameters, and Optimization Parameters. The Upload Dataset step asks the user to upload the real location trace dataset $D$. Once the user chooses an appropriate file for upload, the page shows a progressive loader that displays how much of the file has been uploaded. After the upload is complete, the user moves to the next step (AdaTrace Parameters). In the AdaTrace parameters step, the user is prompted to choose the $Err$ metric to optimize, the privacy budget $\varepsilon$, the cell count for the first level of the adaptive grid, and the number of trials in each iteration of optimization to reduce the inherent randomness caused by DP noise. A screenshot from this step is provided in Figure \ref{fig:inputPhase}. Once these parameters are provided, the user moves to the last step (Optimization Parameters). In this step, the user is permitted to pick the number of random explorations and the number of guided explorations that the Bayesian optimization should take. Larger number of explorations cause optimization to take longer (i.e., longer wait time for the user) but will likely yield better-optimized results. 

\textbf{Computation and Optimization Phase:} In this phase, the user views a page that displays the live-streamed results of the optimization process as it is running in the back-end system (Python component and Java component). A screenshot is provided in Figure \ref{fig:optimizationPhase}. Each step of the Bayesian optimization as well as the corresponding calculated error values are displayed to the user as soon as they are calculated. There is also a progressive loading bar that shows the user how much of the optimization process has been completed thus far. This is to inform the user about how many steps are completed, how many are left to finish, and accordingly, the user can estimate the completion time of optimization. Upon the completion of optimization, a button becomes available at the bottom of this page to take the user to the Results and Analysis phase.

\textbf{Results and Analysis Phase:} In this phase, the user is able to view and analyze the results of OptaTrace, e.g., $D_{syn}$ and related statistics. A screenshot is provided in Figure \ref{fig:resultsPhase}. It can be observed from the screenshot that this phase consists of six tabs on the left hand side. In the ``Input Values" tab, the user can review the values that were chosen in the Input Phase which led to the current results. The ``Results" tab displays the optimized set of parameters found using Bayesian optimization; furthermore, it lists the error values for $D_{syn}$ computed using all of the built-in error metrics from the utility module (see Section \ref{sec:UtilityModule}). This page also enables the user to download $D_{syn}$. The remaining four tabs are for analyzing and visualizing Opta\-Trace's output $D_{syn}$ and comparing it with the original $D$. For example, the screenshot provided in Figure \ref{fig:resultsPhase} is from the ``Spatial Distribution" tab, where the user can see a spatial density heatmap of $D$ and $D_{syn}$ displayed side-by-side. The heatmaps are based on the x-y (or lat, long) coordinates of location traces, divided by 10x10=100 bins. In similar fashion, the ``Trip Distribution" tab visualizes the trip distributions of $D$ and $D_{syn}$ side-by-side, the ``Travel Distance Analysis" tab displays histograms of traces' travel distances in $D$ and $D_{syn}$ side-by-side, etc. The collective goal of these tabs is to provide an early visual insight on the impact of $\varepsilon$-DP on data utility. The tabs are extensible such that new visualizations may be added to the front-end source code. 

\vspace{-4pt}
\section{Experimental Evaluation} \label{sec:Experiments}

\subsection{Experiment Setup} \label{sec:ExpSetup}

\textbf{Datasets:} We experiment with three datasets that were also used in \cite{gursoy2018differentially,gursoy2018utility}. Our first dataset is \textit{Taxi}, which consists of GPS traces of taxis operating in the city of Porto, Portugal. The traces were made available as part of the Taxi Service Prediction Challenge at ECML-PKDD 2015 \cite{moreira2013predicting}. We extracted 15,000 taxi trips from the denser areas in the city to construct our Taxi dataset. Our second dataset is \textit{Brinkhoff-20k}, which contains location traces of vehicles simulated using Brinkhoff's network generator for moving objects \cite{brinkhoff2002framework}. The map of Oldenburg, Germany was used to simulate movements of 20,000 vehicles and their locations were sampled at 15.6 second time intervals. Our third dataset is \textit{Brinkhoff-4k}, which is a small sample consisting of 4,056 traces extracted from the Brinkhoff-20k dataset. The purpose of using both a large version and small version of Brinkhoff is to compare the behavior of errors and optimization on two semantically similar but cardinality-wise different datasets. 

\textbf{Competitors:} We compare OptaTrace with existing work on differentially private location trace synthesis. Our comparison includes three competitors total:
\begin{itemize}
    \item OptaTrace is the system proposed in this paper. When we report a certain type of error for OptaTrace, we assume that the synthesis is optimized for that error metric, e.g., when reporting Query Error for OptaTrace we assume $Err$ = Query Error.
    \item AdaTrace is a state-of-the-art differentially private location trace synthesis system described in \cite{gursoy2018utility}. Results for AdaTrace are reported using the parameter and budget settings used in \cite{gursoy2018utility}.
    \item EQW is a naive version of OptaTrace in which no optimization is performed and OptaTrace is executed with fixed equal weights of $w_1$ = $w_2$ = $w_3$ = $w_4$. EQW is included in the comparison to demonstrate the benefit of OptaTrace's optimization module.
\end{itemize}

\begin{table*}[!ht]
\centering
\caption{Comparing our proposed OptaTrace system against AdaTrace and EQW. Results across four error metrics, three $\varepsilon$ values and three datasets agree that OptaTrace provides higher utility (lower error) compared to AdaTrace and EQW.}
\label{tbl:comparison}
\begin{tabularx}{.98\textwidth}{rr|ccc|ccc|ccc}
\toprule
 &  & \multicolumn{3}{c|}{\textbf{Taxi}} &   \multicolumn{3}{c|}{\textbf{Brinkhoff-4k}} & \multicolumn{3}{c}{\textbf{Brinkhoff-20k}} \\ \midrule
 &  & ~EQW~ & AdaTrace & OptaTrace & ~EQW~ & AdaTrace & OptaTrace & ~EQW~ & AdaTrace & OptaTrace \\ \midrule
 & $\varepsilon=0.5$ & 0.095 & 0.094 & 0.059 & 0.204 & 0.186 & 0.133 & 0.151 & 0.149 & 0.132 \\
\textbf{Query Error} & $\varepsilon=1.0$ & 0.082 & 0.087 & 0.052 & 0.210 & 0.168 & 0.101 & 0.123 & 0.115 & 0.100 \\
 & $\varepsilon=2.0$ & 0.091 & 0.095 & 0.045 & 0.128 & 0.115 & 0.093 & 0.132 & 0.128 & 0.097 \\ \midrule
 & $\varepsilon=0.5$ & 0.509 & 0.481 & 0.418 & 0.571 & 0.549 & 0.518 & 0.491 & 0.480 & 0.458 \\
\textbf{Pat.~Min.~Sup.~Error} & $\varepsilon=1.0$ & 0.460 & 0.429 & 0.359 & 0.503 & 0.474 & 0.445 & 0.419 & 0.408 & 0.375 \\
 & $\varepsilon=2.0$ & 0.391 & 0.378 & 0.339 & 0.465 & 0.485 & 0.423 & 0.414 & 0.407 & 0.379 \\ \midrule
 & $\varepsilon=0.5$ & 0.151 & 0.138 & 0.093 & 0.257 & 0.206 & 0.106 & 0.075 & 0.059 & 0.033 \\
\textbf{Trip Error} & $\varepsilon=1.0$ & 0.096 & 0.074 & 0.027 & 0.162 & 0.097 & 0.058 & 0.036 & 0.019 & 0.011  \\
 & $\varepsilon=2.0$ & 0.025 & 0.019 & 0.009 & 0.098 & 0.076 & 0.035 & 0.018 & 0.015 & 0.008 \\
 \midrule
 & $\varepsilon=0.5$ & 0.038 & 0.038 & 0.025 & 0.099 & 0.091 & 0.069 & 0.063 & 0.064 & 0.055 \\
\textbf{Travel Distance Error} & $\varepsilon=1.0$ & 0.027 & 0.022 & 0.018 & 0.069 & 0.060 & 0.051 & 0.055 & 0.052 & 0.049 \\
 & $\varepsilon=2.0$ & 0.023 & 0.021 & 0.016 & 0.057 & 0.049 & 0.041 & 0.052 & 0.049 & 0.048 \\
\bottomrule
\end{tabularx}
\vspace{-4pt}
\end{table*}

\textbf{Evaluation Metrics:} We use four error metrics in optimization and utility loss measurement: \textit{Query Error}, \textit{Pattern Mining Support Error}, \textit{Trip Error}, and \textit{Travel Distance Error}. According to the utility categories listed in the utility module (Section \ref{sec:UtilityModule}), Query Error falls under the category of Spatial Density and Locality Metrics; Trip Error and Travel Distance Error fall under the category of Spatio-Temporal Travel Metrics; and Pattern Mining Support Error falls under the category of Pattern Mining Metrics. 

\textit{Query Error} is a popular measure for evaluating noisy data quality. Consider spatial counting queries of the form: ``Retrieve the number of traces passing through geographical region X". Let $Q$ denote a query of this form and $Q(D)$ denote its answer when issued on dataset $D$. The Query Error is:
\begin{equation}
    \text{Query Error} = \frac{|Q(D)-Q(D_{syn})|}{max\{Q(D),b\}}
\end{equation}
where $b$ is a sanity bound to mitigate the effect of extremely selective queries. We set $b = 0.01 \times |D|$. We generate 200 random queries by changing the geographical region of the query and report the average Query Error across all queries.

\textit{Pattern Mining Support Error} measures the error in the support values of frequent mobility patterns. Let $P$ denote a pattern as an ordered sequence of cells, e.g., $P: C_3 \rightarrow C_5 \rightarrow C_1$. We define the support of a pattern, $supp(D, P)$, as the number of occurrences of $P$ in dataset $D$. We mine the top-$k$ patterns from the real dataset $D$, i.e., the $k$ patterns with highest support, denoted by $\mathcal{F}_{\mathcal{U}}^k(D)$. Then, the Pattern Mining Support Error is:
\begin{equation}
    \frac{\sum_{P \in \mathcal{F}_{\mathcal{U}}^k(D)} \frac{|supp(D,P)-supp(D_{syn},P)|}{supp(D,P)}}{k}
\end{equation}
We use $k=100$, minimum pattern length of 2 and maximum pattern length of 8 in our experiments. 

\textit{Trip Error} measures error in preserving the correlations between trips' start and end regions. Recall from Section \ref{sec:SynopsisModule} that $\mathcal{R}_{D,\mathbb{A}}$ denotes the trip distribution of dataset $D$ given grid $\mathbb{A}$. We compute the trip distribution of the real dataset using a 6x6 uniform grid $\mathcal{U}$ (denoted $\mathcal{R}_{D,\mathcal{U}}$) and the synthetic dataset using the same grid (denoted $\mathcal{R}_{D_{syn},\mathcal{U}}$). The Trip Error is defined as the Jensen-Shannon divergence between the two distributions: $JSD(\mathcal{R}_{D,\mathcal{U}}, \mathcal{R}_{D_{syn},\mathcal{U}})$.

\textit{Travel Distance Error} measures the aggregate error in trip travel distances (travel lengths). We calculate the total travel distance of a trip by summing the distance between each consecutive location reading in that trip. Upon finding the maximum travel distance from the real dataset $D$, we quantize travel distances into 20 equal sized buckets: $\{ [0,x), [x, 2x), ...,$ $[19x, 20x] \}$, where $20x$ is the longest travel distance present in $D$. For each bucket, we determine how many trips' total travel distances fall into that bucket, thereby obtaining a histogram of travel distance buckets versus counts of trips in each bucket. Let $\mathcal{N}_D$ and $\mathcal{N}_{D_{syn}}$ denote the histograms extracted from $D$ and $D_{syn}$ respectively. Then, the Travel Distance Error is equal to: $JSD(\mathcal{N}_D, \mathcal{N}_{D_{syn}})$. 

\subsection{Comparison with Prior Work} \label{sec:Comparison}

\begin{figure*}
\centering
\begin{minipage}[b]{0.31\textwidth}
\includegraphics[width=\textwidth]{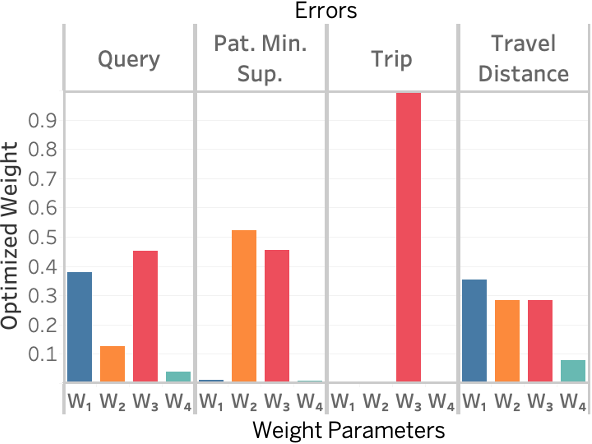}
\subcaption{Brinkhoff-20k $\varepsilon=2$}
\label{fig:optWeightsBrinkhoff20keps2}
\end{minipage}%
\hspace{5mm}
\begin{minipage}[b]{0.31\textwidth}
\includegraphics[width=\textwidth]{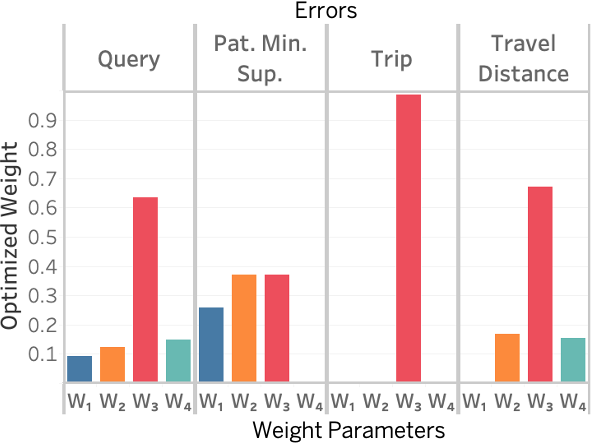}
\subcaption{Brinkhoff-20k $\varepsilon=1$}
\label{fig:optWeightsBrinkhoff20keps1}
\end{minipage}%
\hspace{5mm}
\begin{minipage}[b]{0.31\textwidth}
\includegraphics[width=\textwidth]{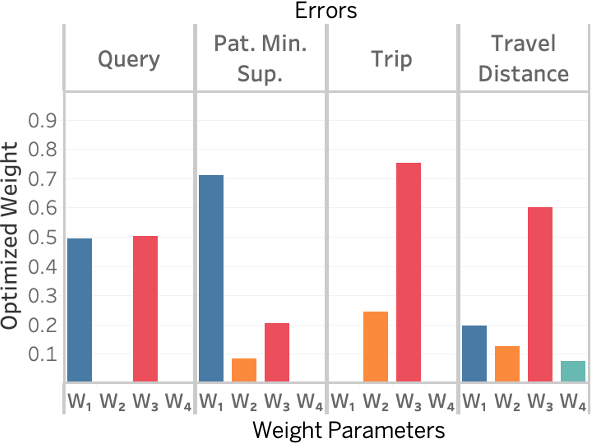}
\subcaption{Taxi $\varepsilon=1$}
\label{fig:optWeightsTaxieps1}
\end{minipage}%
\caption{Optimized values of the weight parameters $w_1$, $w_2$, $w_3$, $w_4$ found using Bayesian optimization in three different dataset and $\varepsilon$ combinations. We observe that the optimized weight values differ from one $Err$ metric to another, as well as from one dataset-$\varepsilon$ combination to another.}
\label{fig:optWeights}
\vspace{-8pt}
\end{figure*}

In Table \ref{tbl:comparison}, we compare OptaTrace with AdaTrace and EQW using four error metrics, three $\varepsilon$ values and three datasets. Results show that OptaTrace's optimized trace synthesis approach yields substantially lower error compared to other approaches. OptaTrace's utility improvement is most pronounced with the Query Error and Trip Error metrics, with roughly 50\% error reduction in terms of Trip Error on average. 

Comparing the results obtained using the Brinkhoff-4k (smaller dataset) versus the Brinkhoff-20k (larger dataset), we observe that OptaTrace beats AdaTrace on both datasets, but the amount of error reduction is different. On the smaller dataset, we observe higher error reduction; whereas on the larger dataset, we observe smaller error reduction. The reason is because in the larger dataset, errors are already relatively lower compared to the smaller dataset. Thus, there is relatively less room for error improvement using Bayesian optimization. As a result, we expect OptaTrace to be useful in reducing error on smaller datasets. We also observe that smaller $\varepsilon$ often yields larger error in all three competitors (EQW, AdaTrace, OptaTrace), as expected. This also brings larger room for error reduction using optimization when $\varepsilon$ is small. Consequently, the difference between OptaTrace's error and EQW's error is often larger when $\varepsilon$ is small (e.g., $\varepsilon =0.5$). Their net difference decreases when $\varepsilon$ is large (e.g., $\varepsilon=2$).

\subsection{Analysis of Weight Parameters}

Having demonstrated the utility improvement of OptaTrace compared to prior work, we now exemplify the need for fresh optimization for each different $D$, $\varepsilon$ and $Err$ metric; rather than using a fixed or pre-defined set of weights across multiple datasets, $\varepsilon$ values or $Err$ metrics. In Figure \ref{fig:optWeights}, we illustrate the optimized weight values $(w_1, w_2, w_3, w_4)$ found using Bayesian optimization for three different dataset and $\varepsilon$ combinations. We make three observations. First, comparing Figure \ref{fig:optWeights}a and \ref{fig:optWeights}b, we observe that under the same dataset and $Err$ metric, the optimized weight values may change depending on the value of $\varepsilon$. Although optimized weight values may be similar for some $Err$ metrics (such as Trip Error), they are significantly different for others (such as Query Error and Travel Distance Error). Second, comparing Figure \ref{fig:optWeights}b and \ref{fig:optWeights}c, we observe that under the same $\varepsilon$ and $Err$ metric, the optimized weight values may also change depending on the dataset, as one can visually observe the differences between the results on Brinkhoff-20k versus Taxi. Finally, Figures \ref{fig:optWeights}a, \ref{fig:optWeights}b and \ref{fig:optWeights}c each individually show that the optimized weight values are different for each different $Err$ metric, when the dataset and $\varepsilon$ values are constant. 

Combining the above three observations, we validate that the optimized weight values depend individually on all three factors: dataset $D$, $\varepsilon$ value, and $Err$ metric. Changing one or more of these factors may result in substantially different optimization results. Consequently, we conclude that OptaTrace's fresh Bayesian optimization for each new dataset, $\varepsilon$, and $Err$ metric is beneficial in improving utility; rather than re-using weight values that were learned under different conditions. 

\subsection{Optimization Process and Convergence}

\begin{figure}
    \centering
    \includegraphics[width=0.45\textwidth]{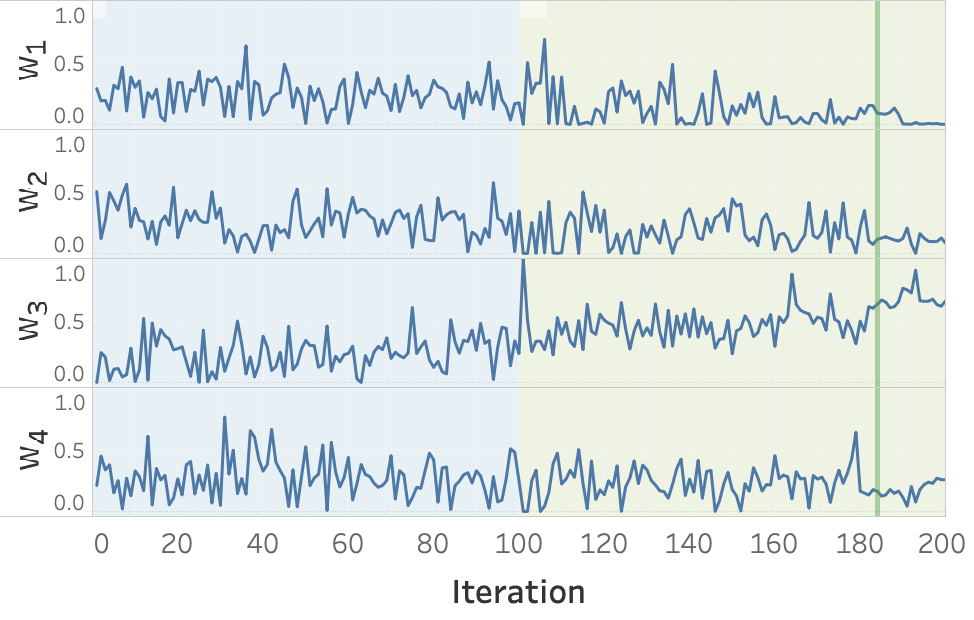} 
    \vspace{-2pt}
    \caption{Values of weight parameters versus iterations of Bayesian optimization (100 iterations of search space exploration + 100 iterations of optimization).}
    \label{fig:optimizationProcess}
    \vspace{-6pt}
\end{figure}

In Figure \ref{fig:optimizationProcess}, we provide a sample run of the Bayesian optimization process used in OptaTrace. We initiate optimization with the Brinkhoff-4k dataset, $\varepsilon=1$ and $Err$ = Travel Distance Error, with 100 iterations of search space exploration and 100 iterations of optimization. We track the values of each of the weight parameters $w_1, w_2, w_3, w_4$ for the whole 200 iterations. The resulting graph is illustrated in Figure \ref{fig:optimizationProcess}. 

The graph shows that during the search space exploration phase (first 100 iterations) and the early stages of the Bayesian optimization phase (iterations 100-140), there can be larger variances in weight values, as indicated by the large spikes and drops between consecutive iterations. However, as optimization approaches the later rounds, most of the weight values converge to their optimized values, and their variances become smaller. Particularly, in iterations between 180-200, each of the weight values become stable, e.g., $w_1$ converges to a small non-zero value, $w_3$ converges to a value closer to 1, and $w_2$ and $w_4$ converge to their optimized values between 0 and 0.5. We conclude from this example that reasonably stable convergence can be reached within 200 iterations. 

\section{Related Work} \label{sec:RelatedWork}
 
Differentially private data synthesis and publication have been active areas of research over the last decade. Several methods were developed for tabular data \cite{mohammed2011differentially,xu2017dppro,zhang2017privbayes}, set-valued data \cite{chen2011publishing,zhang2013differentially}, sequential data \cite{chen2012differentially}, transit data \cite{chen2012stm}, and location traces \cite{he2015dpt,gursoy2018utility,gursoy2018differentially,wang2020protecting,deldar2020enhancing,ding2020differentially}. Among them, the methods on differentially private location trace synthesis (DPLTS) are most relevant to our work. In this domain, He et al.~developed the DPT system for private trajectory synthesis using hierarchical reference systems, i.e., location discretization with hierarchically organized grids \cite{he2015dpt}. SGLT system was developed in \cite{bindschaedler2016synthesizing}, which synthesizes location traces that satisfy a privacy notion called \textit{plausible deniability}. Plausible deniability relies on a privacy test to reject a candidate synthetic trace from being added to $D_{syn}$ if there are not enough ``similar" traces to it in $D$. It was later shown in \cite{bindschaedler2017plausible} that a randomized version of this test can yield a restricted form of $(\varepsilon,\delta)$-DP for a certain set of $(\varepsilon,\delta)$ parameters. More recently, Gursoy et al.~developed DP-Star which was later superseded by the AdaTrace system \cite{gursoy2018differentially,gursoy2018utility}. In \cite{gursoy2018utility}, AdaTrace was compared to prior works such as DPT and SGLT, and it was shown that AdaTrace provides superior utility overall. Therefore, we compare our proposed OptaTrace system mainly with AdaTrace. 

\vspace{-1pt}
\section{Conclusion} \label{sec:Conclusion}
\vspace{-1pt}

We presented OptaTrace, an optimization-based approach to differentially private location trace synthesis. Given a real trace dataset $D$, privacy budget $\varepsilon$ and the $Err$ metric, OptaTrace uses Bayesian optimization to minimize $Err(D, D_{syn})$ while ensuring $D_{syn}$ satisfies $\varepsilon$-differential privacy. Compared to prior work, our optimization-based approach is shown to provide substantial error reduction and utility improvement. Contributions of OptaTrace also include a utility module for convenient error measurement and $Err$ metric selection; and a front-end web interface for accessible and easy-to-use DPLTS functionality for non-experts.


\vspace{-2pt}
\section*{Acknowledgments}
\vspace{-2pt}

The authors acknowledge partial support by the National Science Foundation under grants NSF 2038029, NSF 1564097, and an IBM faculty award.

\vspace{-2pt}

\bibliographystyle{IEEEtran}
\bibliography{references.bib}

\end{document}